\documentstyle[12pt]{article}
\begin{document}
\renewcommand{\thefootnote}{\fnsymbol{footnote}}
\newpage
\pagestyle{empty}
\setcounter{page}{0}

\vfill
\begin{center}

{\LARGE {\bf {\sf
Nonlinear transforms of momenta and Planck scale limit
}}} \\[0.8cm]
{\large A.Chakrabarti

{\em

Centre de Physique Th\'eorique\footnote{Laboratoire Propre
du CNRS UPR A.0014}, Ecole Polytechnique, 91128 Palaiseau Cedex, France.\\
e-mail chakra@cpht.polytechnique.fr}}

\end{center}

\smallskip

\smallskip

\smallskip

\smallskip

\smallskip

\smallskip

\begin{abstract}
Starting with the generators of the Poincar\'e group for arbitrary
mass (m)
and spin (s) a nonunitary transformation is implemented to obtain momenta
with an absolute Planck scale limit. In the rest frame (for $m>0$) the
transformed energy coincides with the standard one, both being $m$.
As the latter tends to infinity under Lorentz transformations the
former tends to a finite upper limit $m\coth(lm) = l^{-1}+ O(l)$
where $l$ is the Planck length and the mass-dependent nonleading
terms vanish exactly for zero rest mass.The invariant $m^{2}$ is
conserved for the transformed momenta. The speed of light continues
to be the absolute scale for velocities. We study various aspects of
the kinematics in which two absolute scales have been introduced in
this specific fashion. Precession of polarization and transformed
position operators are among them. A deformation of the Poincar\'e
algebra to the $SO(4,1)$ deSitter one permits the implementation of
our transformation in the latter case. A supersymmetric extension of
the Poincar\'e algebra is also studied in this context.
\end{abstract}

\vfill

\newpage

\pagestyle{plain}

\section{Introduction}
  Possible modifications of special relativity introducing, in addition to
the velocity of light, a second invariant scale corresponding to the Planck
energy ( the inverse of the Planck length ) have been studied in numerous
recent papers exploring various aspects $[1,2,3,4,5,6,7,8,9,10,11]$. The
titles of these papers ( citing other relevant sources ) convey some idea of
the topics addressed. Among the papers cited above our work can be compared
most directly, concerning both analogies and crucial differences, with the
work of Magueijo and Smolin $[4,5]$.Like them we introduce the nonlinear
constructions via a nonunitary transformation. But (unlike all the foregoing
studies) we introduce spin at the outset. To be able to do so adequately
we start with an irreducible representation $[m,s]$ of the Poincar\'e
group of positive rest mass $(m>0)$ and an arbitrary integer or
half-integer spin $s$. The momentum generators are thus constrained to
satisfy

\begin{equation}
P^{\mu}P_{\mu} = {P^2}_{0} -{\overrightarrow{P}}^2 = m^2
\end{equation}
 It will be implicit henceforth that (with positive squareroot and
${\bigtriangledown}_i$ denoting derivative with respect to $P_i$ )

\begin{equation}
P_0 = \sqrt{{\overrightarrow{P}}^2 + m^2},\qquad
\overrightarrow{\bigtriangledown}{P_0} = \frac{\overrightarrow{P}}{P_0}
\end{equation}

Introducing $(2s+1)\times (2s+1)$ spin matrices $\overrightarrow{S}$
satisfying
\begin{equation}
 [S_i ,  S_j] = i{\epsilon}_{ijk}S_k
\end{equation}
the generators of pure rotations $(\overrightarrow{J})$ and those
of pure Lorentz transformations  $(\overrightarrow{K})$ can be
represented as
\begin{equation}
  \overrightarrow{J}= -i\overrightarrow{P}\times
\overrightarrow{\bigtriangledown} + \overrightarrow{S}
\end{equation}

\begin{equation}
  \overrightarrow{K}= -iP_0\overrightarrow{\bigtriangledown} -
\frac{\overrightarrow{P}\times\overrightarrow{S}}{P_0+m}
\end{equation}

  Let us briefly note the following points $[12]$ for later use.

   (a): The first term of $\overrightarrow{K}$ should {\it not} be
symmetrized. The hermiticity of $\overrightarrow{K}$ and the relation to
Newton-Wigner position operators are discussed in Ref.$12$ ( from eqn.
$(2.18)$ onwards ).

  (b): For $m=0$, the last term of $\overrightarrow{K}$ is not
well-defined only for energy-momenta $(0,0,0,0)$, namely at the tip of the
light cone which is never in the same orbit with
         $$   p^2=0,\qquad p_0 \neq 0 $$

Hence excluding massless particles with strictly zero energy one can
consistently use $(5)$ with $m=0$. We will present below results for $m=0$
obtained systematically in this fashion. An explicit unitary transformation
to Wigner's construction in terms of the little group $E_2$ has been
presented elsewhere ( see the discussion and the references in Ref.$12$
from eqn.$(2.21)$ onwards ). There it is explained how , inspite of the
presence of three spin components in $(4)$ and $(5)$ one finally deals with
only one conserved helicity component for $m=0$.

  (c): The canonical form given by $(4)$ and $(5)$ is valid for any spin.
The Pauli-Lubansky $4$-vector is obtained by contracting the dual of the
tensor $(\overrightarrow{J}, \overrightarrow{K})$ by $P_{\mu}$ or
equivalently as

\begin {equation}
W_{\mu}=-i[P_{\mu},\overrightarrow{K}.\overrightarrow{J}]
\end{equation}
\begin {equation}
=(\overrightarrow{P}.\overrightarrow{J},P_0 \overrightarrow{J} -
\overrightarrow{P}\times\overrightarrow{K})
\end{equation}

\begin {equation}
=(\overrightarrow{P}.\overrightarrow{S},m \overrightarrow{S} +
(P_0 +m)^{-1}(\overrightarrow{P}.\overrightarrow{S})\overrightarrow{P})
\end{equation}
  This satisfies

\begin {equation}
W_{\mu}W^{\mu}=-m^2{\overrightarrow{S}}^2
\end{equation}
 The relation , for $s=1/2$, with the the Dirac representation and the Dirac
equation are indicated in Ref.$12$ (from eqn.$(2.49)$ onwards). The
relevant transformation relating the two representations diagonalizes the
Dirac mass operator $(\gamma.p) $.

  Before introducing the explicit form of the transformation $V$ (to be
presented below) let us note the following aspect.

  Having explicitly constructed all the transforms

\begin {equation}
V(P_0,\overrightarrow{P},\overrightarrow{J}, \overrightarrow{K})V^{-1}
\end{equation}
one can study the set

 $$V(\overrightarrow{J},\overrightarrow{K})V^{-1},(P_0,\overrightarrow{P})$$

or, in a complimentary fashion, the set

$$(\overrightarrow{J},\overrightarrow{K}),V(P_0,\overrightarrow{P})V^{-1}$$

 In the latter case one conserves the explicit representation of the
Lorentz algebra and associates physical significance with the
transformed momenta. We will adopt the latter approach below (
providing however  complete results for $(10)$ ). This will furnish
, in the terminology of Ref.7, an example of $DSR2$ theories with
bounded energy and momenta. The inverse formulae, giving the
standard momenta in terms of the transformed ones are obtained very
simply.

\section{The Transformation:}

 Let

\begin{equation}
V= \exp(-lP_{0}\overrightarrow{P}.\overrightarrow{\bigtriangledown})
\end{equation}
where
      $0<l\ll {1}$,  and    $ P_0 = \sqrt {{\overrightarrow{P}}^2 +m^2}$.

 In fact, one may assume that in our chosen units $(c=1)$ $l$ is the Planck
length which is more generally

\begin{equation}
l_P =\sqrt{(\bar {h}G/{c^3})}
\end{equation}

  As compared to the corresponding operator

\begin{equation}
U^{-1}=exp(-l_{P}p_{0}p^{\mu}{\partial}_\mu)
\end{equation}
of $Ref.4$, we have kept only the three space components in the
scalar product but still with the factor $P_0$ rather than
$P=\mid{\overrightarrow{P}}\mid$. This is crucial for the remarkable
properties obtained below for arbitrary spin.

  Implementing, consistently with $(2)$, for any $f$,

$$[\overrightarrow{\bigtriangledown},{P_0}]f=\frac{\overrightarrow{P}}{P_0}f$$
one obtains
\begin{equation}
\overrightarrow{\textsf{P}}\equiv
V{\overrightarrow{P}}V^{-1}=\frac{m\overrightarrow{P}}{sh(lm) P_0+  ch(lm)m}
\end{equation}

\begin{equation}
\textsf{P}_0\equiv
VP_0{V^{-1}}=m\frac{ch(lm) P_0+  sh(lm)m}{sh(lm) P_0+  ch(lm)m}
\end{equation}

satisfying

\begin{equation}
{\textsf{P}_0}^2 -{\overrightarrow{\textsf{P}}}^2 =
{P_0}^2-{\overrightarrow{P}}^2 =m^2
\end{equation}

\begin{equation}
V \overrightarrow{J}V^{-1}=  \overrightarrow{J}= -i\overrightarrow{P}\times
\overrightarrow{\bigtriangledown} + \overrightarrow{S}
\end{equation}

\begin{equation}
 V \overrightarrow{K}V^{-1}=
-ich(lm) P_0\overrightarrow{\bigtriangledown} -ish(lm)
(m\overrightarrow{\bigtriangledown}+m^{-1}\overrightarrow{P}
(\overrightarrow{P}.\overrightarrow{\bigtriangledown}))
-e^{-lm}\frac{\overrightarrow{P}\times\overrightarrow{S}}{P_0+m}
\end{equation}

  Note the simplicity of the spin dependent part on the right of $(18)$.
This corresponds to

\begin{equation}
V\frac{\overrightarrow{P}}{P_0+m} V^{-1}=
e^{-lm}\frac{\overrightarrow{P}}{P_0+m}
\end{equation}

Hence, squaring each side and using $(1)$ one obtains

\begin{equation}
V\frac{P_0 -m}{P_0+m} V^{-1}=e^{-2lm}\frac{P_0 -m}{P_0+m}
\end{equation}

 Thus one readily obtains $(15)$ and hence also $(14)$. If, without knowing
$(15)$ beforehand, one proceeds directly to compute the power series

\begin{equation}
\textsf{P}_0 = VP_{0}V^{-1}=P_0
-l[P_{0}(\overrightarrow{P}.\overrightarrow{\bigtriangledown}),P_0]
+l^2[P_{0}(\overrightarrow{P}.\overrightarrow{\bigtriangledown})
[P_{0}(\overrightarrow{P}.\overrightarrow{\bigtriangledown}),P_0]]-...
\end{equation}
one obtains
\begin{equation}
\textsf{P}_0 = P_0 -l ({P_0}^2 - m^2) +l^2 P_0 ({P_0}^2 - m^2) - ..
\end{equation}

The series is difficult to sum up. On the other hand,developing
$(15)$ in powers of $l$ one  easily reproduces $(22)$. One similarly
obtains for the modulus of the momentum

\begin{equation}
\textsf{P}=P - lPP_0 + {\frac{1}{2}} {l^2}P({P_0}^2 +P^2) -...
\end{equation}

 It is easy to verify that $(14)$ and $(15)$ can be invertd by simply
changing the sign of $l$. Thus

\begin{equation}
P_0=m\frac{ch(lm)\textsf{P}_0 -  sh(lm)m}{-sh(lm)\textsf{P}_0+  ch(lm)m}
\end{equation}

 and so on. This is consistent with the invariance of
$(P_{0}(\overrightarrow{P}.\overrightarrow{\bigtriangledown}))$ under the
transformation.

  Let us now consider momentum eigenstates and denote the eigenvalues of
$P_{\mu}$ and $\textsf{P}_{\mu}$ by $p_{\mu}$ and $p'_{\mu}$
respectively. Then
\begin{equation}
 p'_0 = m \frac{coth(lm) p_0 +m}{p_0 +coth(lm)m}
\end{equation}
Hence,since we are considering positive $p_0$,$m$ and $l$,

  $$p'_0 < coth(lm) m$$

Similarly for the modulus of the momentum one obtains

  $$p' <\frac{m}{sh(lm)}$$

 {\it Thus our transformation, valid for arbitrary spin, indeed leads to
an invariant energy scale }. This is the crucial property. For

$$ p_0 = m,\qquad p'_0 =m$$

 and as
$$p_0 \rightarrow \infty, \qquad p'_0 \rightarrow  coth(lm) m$$

from below. {\it For all observers }$p'_0$ {\it remains bounded}. Starting
together with $p_0$ in the rest frame $p'_0$ lags progressively behind as
the former increases to finally encounter the barrier $mcoth(lm)$. In
powers of $l$ one obtains

\begin{equation}
p'_0 < l^{-1} +\frac{1}{3}m^{2}l +O(l^2)
\end{equation}
and
\begin{equation}
p' < l^{-1} -\frac{1}{6}m^{2}l +O(l^2)
\end{equation}

 The modulus of the transformed velocity, quite consistently with our
chosen units $(c=1)$, has the high energy limit ,for $p_0 \rightarrow
\infty$, given by

\begin{equation}
 \frac{p'}{p'_0} \rightarrow \frac{1}{ch(lm)} = 1 - \frac{1}{2}l^2m^2 <1
\end{equation}

The limit $1$ is attained exactly for for $m=0$. this will be seen more
precisely immediately below.

\section{Zero rest mass:}

As explained in note $(b)$ following eqn.$(5)$,,the essential results for
$m=0$ can be obtained (rather than starting again with $m=0$ in $V$ )
easily and directly from our previous ones.Thus for
$m\rightarrow 0$ one obtains from $(14)$ and $(15)$

\begin{equation}
\overrightarrow {\textsf{P}}= \frac{\overrightarrow P}{lP_0 +1}
\end{equation}
and
\begin{equation}
\textsf{P}_0 = \frac{P_0}{lP_0 +1}
\end{equation}
satisfying evidently (like $P_0$ and $\overrightarrow P$ )

$${\textsf{P}_0}^2 - {\overrightarrow {\textsf{P}}}^2 = 0  $$

 Now as compared to the inequality following $(25)$, again for all
parameters positive,

\begin{equation}
 p'_0 =\frac{p_0}{lp_0 +1} <l^{-1}
\end{equation}
As
 $$p_0 \rightarrow \infty,\qquad p'_0 \rightarrow l^{-1}$$
from below. And as compared to $(28)$,
\begin{equation}
 \frac{p'}{p'_0} = \frac{p}{p_0} = 1
\end{equation}

Thus, considering all masses and the system

$$(\overrightarrow{J},\overrightarrow{K},
\textsf{P}_0,\overrightarrow{\textsf{P}})$$

one indeed implements {\it two absolute scales, one for
velocity}$(c=1)$ {\it and one for energy} . The leading term for the
limiting energy is always $l^{-1}$. This becomes exact for zero rest
mass. For positive mass the exact result is provided by $(25)$.

\section{Precession of polarization:}

  Since $V$ commutes with $\overrightarrow S$ the standard results for
precession of polarization are conserved. (See the complete discussion in
Ref.$12$.) They can however be reexpressed in terms of
$(\textsf{P}_0,\overrightarrow{\textsf{P}})$ if so desired. Thus under an
infinitesimal Lorentz transformation of velocity $tanh\chi$
$(\rightarrow \chi )$ parallel to the unit vector $\hat n$, the
change

\begin{eqnarray}
\delta \overrightarrow S &=&
i[\chi\hat{n}.\overrightarrow{K},\overrightarrow S]\nonumber\\
&=&-\chi\frac{(\hat{n}\times\overrightarrow{P})\times\overrightarrow{S}}{P_0+m}\
\nonumber\\
&=&-\chi{e}^{lm}\frac{(\hat{n}\times\overrightarrow{\textsf
{P}})\times\overrightarrow{S}}{\textsf {P}_0+m}
\end{eqnarray}

 Thus the formal expression is altered by a simple overall factor $e^{lm}( =
1+O(l))$. In Ref.$12$ it is explained how $(34)$ leads to the famous Thomas
factor $\frac{1}{2}$. We will not go further into such topics in the
present study.

          We indicate below very briefly possible generalizations of our
study in two different directions.

\section{Deformation of Poincar\'e to $SO(4,1)$ deSitter algebra:}

  The Lorentz algebra has two invariants,
$$ (\overrightarrow K.\overrightarrow J), \qquad ({\overrightarrow
{K}}^2 -{\overrightarrow {J}}^2 )$$

 As pointed out before ( see eqns.$(6)$ to $(9)$ ), commuting
$P_{\mu}$ with the first one leads to $W_{\mu}$ giving the spin.
Commutation of
$P_{\mu}$ with the second leads to the homogenous $SO(4,1)$ algebra where (
along with the Lorentz $SO(3,1)$ generators and $\mu =(0,1,2,3)$ )

\begin{equation}
L_{\mu 5} = \frac{i}{M} [({\overrightarrow
{K}}^2 -{\overrightarrow {J}}^2 ), P_{\mu}] +\lambda P_{\mu}
\end{equation}
 Here $M={(P^{\mu}P_{\mu})}^{\frac{1}{2}}$ is the mass operator and
$\lambda$ is an arbitrary parameter.Starting with an irreducible space
$[m,s]$ ( with $m>0$,say ) one can compute explicitly the actions of
$L_{\mu 5}$ on the states using $(36)$. Elsewhere $[13]$ we have studied
$(36)$ in a more general context using however the Lorentz basis. Here we
only point out that $( VL_{\mu 5}V^{-1})$ is obtained directly from our
foregoing results. A detailed study is beyond the scope of this paper.

\section {A supersymmetric extension:}

  A simple supersymmetric extension $[14]$ permitting a ready implementation
of our transformation can be obtained as follows.(Previous sources are
cited in $[14]$.) One starts with two fermionic operators satisfying
(for $ i = 1,2 $)

 $$[a_i,a_j]_{+} =0=[{a^\dagger}_i,{a^\dagger}_j]_{+}, \qquad
[a_i,{a^\dagger}_j]_{+} =\delta_{ij}$$

 One defines $Q=(Q_1,Q_2)$ and the adjoint $Q^\dagger$ ( a column
with two rows ) as

\begin{equation}
Q^\dagger = \frac{1}{\sqrt{2(P_0 +M)}}[(P_0 +M)
+\overrightarrow{\tau}.\overrightarrow{P}] a^\dagger
\end{equation}

 Then in terms of Pauli matrices

\begin{equation}
 [Q^\dagger,Q]_+ = {\tau}_0 P_0 +
\overrightarrow{\tau}.\overrightarrow{P}
\end{equation}

 This compact notation implies symmetrization of each term of
the $2\times{2}$ matrix $ {Q^\dagger} Q$. Thus, for example, at
the top right one obtains
$Q^{\dagger}_{1} Q_{2} + Q_{2} Q^{\dagger}_{1} = P_{1} -iP_{2}$

 A Majorana spinor is provided by $(Q_1,Q_2,{Q^\dagger}_2,-{Q^\dagger}_1)$.

  Next one defines

\begin{equation}
\overrightarrow \Sigma = \frac{1}{2} (a \overrightarrow{\tau} a^\dagger)
\end{equation}

  Adding the spin operator $\overrightarrow \Sigma $ to $\overrightarrow S$
define

\begin{equation}
  \overrightarrow{J}= -i\overrightarrow{P}\times
\overrightarrow{\bigtriangledown} + (\overrightarrow{S} +\overrightarrow
\Sigma )
\end{equation}

\begin{equation}
  \overrightarrow{K}= -iP_0\overrightarrow{\bigtriangledown} -
\frac{\overrightarrow{P}\times(\overrightarrow{S}+\overrightarrow
\Sigma)}{P_0+m}
\end{equation}

 Now $(P_{\mu},\overrightarrow{K},\overrightarrow{J})$ continue to satisfy
the Poincar\'e algebra along with

\begin{equation}
[ \overrightarrow{J},Q^\dagger ] = - \frac{1}{2}
\overrightarrow{\tau}Q^\dagger
\end{equation}

\begin{equation}
[ \overrightarrow{K},Q^\dagger ] = - \frac{i}{2}
\overrightarrow{\tau}Q^\dagger
\end{equation}

 Thus $(38),(42),(43)$ together complete the supersymmetric extension.
Various aspects are studied in Ref.$14$ citing other sources. Here we only
note that $\overrightarrow\Sigma $ commutes with $V$ and denoting

$$ \tilde{Q}=VQV^{-1}$$

\begin{equation}
\tilde{Q}^\dagger = \frac{1}{\sqrt{2(\textsf{P}^0
+M)}}[(\textsf{P}_0+M)
+\overrightarrow{\tau}.\overrightarrow{\textsf{P}}] a^\dagger
\end{equation}
and
\begin{equation}
 [{\tilde{Q}^\dagger,\tilde{Q}}]_+ = {\tau}_0
\textsf{P}_0 +
\overrightarrow{\tau}.\overrightarrow{\textsf{P}}
\end{equation}

Thus our transformation can be readily implemented for such an extension.
A more detailed study is beyond the scope of this paper.

\section{Gradient operators for $\protect\overrightarrow{\textsf{P}}$:}

  One obtains for transforms of
$\overrightarrow{\bigtriangledown}$, consistently with $(15)$and$(18)$ and
with $\textsf{P}_0$ given by $(15)$,

\begin{equation}
  \overrightarrow{\xi}= \frac{1}{\textsf{P}_0}\biggl((ch(lm)P_0
+sh(lm)m)\overrightarrow{\bigtriangledown} +m^{-1}sh(lm) \overrightarrow{P}
(\overrightarrow{P}.\overrightarrow{\bigtriangledown})\biggr)
\end{equation}
  where

$$  \overrightarrow{\xi}
\equiv V \overrightarrow{\bigtriangledown}V^{-1}$$
  Hence, for such ${\xi}_i$ ,
\begin{equation}
[{\xi}_i,\textsf{P}_j] = {\delta}_{ij}
\end{equation}
and
\begin{equation}
[{\xi}_i,{\xi}_j] = 0
\end{equation}

 Substituting for ${\xi}_i$

$${\xi '}_i={\xi}_i +Vf_iV^{-1}$$

 where $f_i$ depends only on the momenta conserves $(47)$ but not
necessarily $(48)$ unless
$$({\partial}_i f_j -{\partial}_j f_i) =0$$

 In particular, starting with the localizing and hermitian Newton-Wigner
position operators ( Ref.12 from eqn.(2.18) onwards ), namely,

\begin{equation}
\overrightarrow{X} = i\overrightarrow{\bigtriangledown} -
\frac{\overrightarrow{P}}{2{P_0}^2}
\end{equation}
one obtains
\begin{equation}
V\overrightarrow{X} V^{-1}=
i\overrightarrow{\xi} -\frac{\overrightarrow{\textsf{P}}}{2{\textsf{P}_0}^2}
\end{equation}

The components continue to satisfy $(47)$ and $(48)$ ( with a factor $i$ ).

  We will not attempt to explore here whether other choices can lead
to interesting noncommutative Hopf algebras for the coordinates. Both
noncommutative ( Ref.$9$  and sources cited ) and commutative [5]
space-times have been proposed in the context of Planck scale limits of
momenta. In our formalism, apart from the commutativity of $(48)$ ,
the time $t$ remains a parameter ( $P_0$ being given by $(2)$ ).

\section{Conclusion:}

  For all mass and spin we have obtained nonlinear functions of the
standard momenta possessing a Planck scale limit. Our construction
exhibits that such a property is quite consistent with a fixed velocity of
light, time remaining a parameter and commuting position operators
corresponding to those for the nonlinear momenta. Even if one
deliberately seeks a different formalism violating such properties,
comparison with our formalism will provide a deeper understanding.

 Due to the fact that the new momenta are introduced via a
relatively simple conjugation, by our $V$, all relevant properties
are obtained fairly easily and systematically. This has permitted a
ready passage to deSitter $SO(4,1)$ and to a supersymmetric
extension as well.

    Elsewhere $[15]$ we have presented explicit constructions for the
genarators of the Poincar\'e group for spacelike momenta and for lightlike
momenta with continuous spin. We just mention that they have strong
analogies with those introduced here for the timelike case and thus
may suggest how our transformation can be adapted to those cases.

\smallskip

\smallskip

\bibliographystyle{amsplain}

\end{document}